\documentclass[12pt]{article}
\usepackage{cite}
\usepackage{color}
\usepackage{graphicx}
\usepackage{amsmath}
\usepackage{amssymb}
\usepackage{xspace}

\def\baselinestretch{1.2}
\newcommand{\be}{\begin{equation}}
\newcommand{\ee}{\end{equation}}
\newcommand{\beq}{\begin{eqnarray}}
\newcommand{\eeq}{\end{eqnarray}}

\newcommand{\gone}[1]{{}}

\begin{document}
\begin{titlepage}
\begin{flushright}
MAD-TH-08-02
\end{flushright}
%\vspace{12 mm}

\vfil\
%vfil

\begin{center}

{\Large{\bf Melvin Twists of global $AdS_5 \times S_5$ and their Non-Commutative Field Theory Dual}}

\vfil

Danny Dhokarh, Sheikh Shajidul Haque, and Akikazu Hashimoto

\vfil

Department of Physics, University of Wisconsin, Madison, WI 53706

\vfil

\end{center}

%%%%%%%%%%%%%%%%%%%%%%%%%%%%%%%%%%%%%%%%%%%%%%%%%%%%%%%%%%%%%%%%%%%%%%%%%%%%%%%%%%%%%%%
\begin{abstract}
\noindent We consider the Melvin Twist of $AdS_5 \times S_5$ under
$U(1) \times U(1)$ isometry of the boundary $S_3$ of the global
$AdS_5$ geometry and identify its field theory dual. We also study the
thermodynamics of the Melvin deformed theory.
\end{abstract}
%%%%%%%%%%%%%%%%%%%%%%%%%%%%%%%%%%%%%%%%%%%%%%%%%%%%%%%%%%%%%%%%%%%%%%%%%%%%%%%%%%%%%%%%%
\vspace{0.5in}

\end{titlepage}
\renewcommand{\baselinestretch}{1.05}  %Line spacing
%%%%%%%%%%%%%%%%%%%%%%%%%%%%%%%%%%%%%%%%%%%%%%%%%%%%%%%%%%%%%%%%%%%%%%%%%%%%%%%%%%%%%%%%%%%%%

Melvin twist, also known as the T-s-T transformation, is a powerful
solution generating technique in supergravity and string theories
\cite{Dowker:1993bt,Dowker:1994up,Behrndt:1995si,Costa:2000nw,Gutperle:2001mb,Costa:2001ifa}. The procedure
relies on having a $U(1) \times U(1)$ compact isometry along which one
performs a sequence of T-duality, twist, and a T-duality. The twist is
an $SL(2,R)$ transformation on the complex structure of the T-dual
torus. As such, the Melvin twist can simply be thought of as an
$SL(2,R)$ transformation acting on the K\"ahler structure of the torus
parameterized by $U(1) \times U(1)$.

Interesting closed string backgrounds, such as Melvin universes, null
branes, pp-waves, and G\"odel universes can be constructed by applying
the Melvin Twist procedure to the Minkowski background. The
construction reveals the hidden simplicity of these closed string
backgrounds: they are dual to flat spaces. As a result, world sheet
sigma model for strings in these backgrounds are exactly solvable and
have been studied extensively
\cite{Russo:1994cv,Russo:1995tj,Russo:1995aj,Russo:1995ik,Tseytlin:1994ei,Tseytlin:1995fh}.
The same procedure can be applied to black $p$-brane backgrounds to
construct various asymptotically non-trivial space-time geometries
\cite{Gimon:2003xk}.

Melvin twist applied to the $Dp$-brane background and the subsequent
near horizon limit gives rise to supergravity duals for a variety of
decoupled field theories\footnote{An earlier discussion of a construction of this type is \cite{Costa:2001ifa}.} depending on the orientation of the brane and
the Melvin twist.  If both of the $U(1)$ isometries are along the
brane, one generally obtains a non-commutative field theory, typically
with non-constant non-commutativity parameter
\cite{Hashimoto:1999ut,Aharony:2000gz,Hashimoto:2002nr,Dolan:2002px,Hashimoto:2004pb,Hashimoto:2005hy}. If
one of the $U(1)$ is transverse to the brane, then one obtains a
dipole field theory
\cite{Bergman:2000cw,Bergman:2001rw,Ganor:2002ju}. Taking both of the
$U(1)$'s to be transverse to the brane gives rise to the construction
of Lunin and Maldacena \cite{Lunin:2005jy}.  The list of models
constructed along these lines is summarized in table
\ref{table1}. These theories are S-dual to NCOS theories
\cite{Cai:2002sv,Cai:2006tda}. They are also closely related to ``Puff
Field Theory'' which was studied recently in
\cite{Ganor:2006ub,Ganor:2007qh}. The hidden simplicity of Melvin
twists in the context of gauge theory duals manifests itself as
preservation of integrability.  The fact that $q/\beta$-deformed
${\cal N}=4$ SYM remains integrable was pointed out in
\cite{Roiban:2003dw,Berenstein:2004ys}. A broader class of integrable
twists were studied in \cite{Beisert:2005if,McLoughlin:2006cg}.

\begin{table}
\centerline{\begin{tabular}{|l|l|} \hline
Type of Twist & Model \\ \hline \hline 
Melvin Twist & Hashimoto-Thomas model
 \\ \hline
Melvin Shift Twist & Seiberg-Witten Model\\ \hline
Null Melvin Shift Twist & Aharony-Gomis-Mehen model\\\hline
Null Melvin Twist & Dolan-Nappi model \\ \hline
Melvin Null Twist & Hashimoto-Sethi model\\ \hline
Melvin R Twist & Bergman-Ganor model 
\\ \hline
Null Melvin R Twist & Ganor-Varadarajan model \\ \hline
R Melvin R Twist & Lunin-Maldacena model
 \\ \hline
\end{tabular}}
\caption{Catalog of non-commutative gauge theories viewed as a world
volume theory of D-branes in a ``X'' Melvin ``Y'' twist
background. This table originally appeared in
\cite{Hashimoto:2004pb}. \label{table1}}
\end{table}

In this article, we consider the effect of twisting along the $U(1)
\times U(1) \in SO(4)$ isometry of the $S_3$. More specifically, we
consider $AdS_5 \times S_5$ solution of type IIB theory
\beq ds^2& =&  R^2\left[ - \cosh^2 \rho d \tau^2  + d \rho^2 + \sinh^2 \rho (d \theta^2 + \sin^2\theta d \phi_1^2 +\cos^2 \theta d \phi_2^2) 
+ d \Omega_5^2 \right]\cr
B & = & 0\cr
e^\phi & = & {\lambda \over 4 \pi N} \label{ads5xs5}
\eeq
where $\lambda$ is the 't Hooft coupling 
\be \lambda = 2 g_{YM}^2 N = 4 \pi g_s N  = {R^4 \over \alpha'^2}\ , \ee
and perform a Melvin twist on the torus parameterized by the
coordinates $(\phi_1,\phi_2)$. This is equivalent to acting on the
Kahler structure
\be \rho = {1 \over \alpha'} \left(B_{\phi_1 \phi_2} + i \sqrt{g_{\phi_1 \phi_1} g_{\phi_2 \phi_2}}\right) \ee
by an $SL(2,R)$ transformation
\be \rho \rightarrow  \rho' = {\rho \over \chi \rho + 1} \ee
giving rise to a background
\beq ds^2 &=&   \alpha' \sqrt{\lambda} \left[ - \cosh^2 \rho d \tau^2  + d \rho^2 + \sinh^2 \rho \left(d \theta^2 + {\sin^2\theta d \phi_1^2 +\cos^2 \theta d \phi_2^2 \over 1 + \chi^2 \lambda \cos^2 \theta \sin^2 \theta \sinh^4 \rho}\right) 
+ d \Omega_5^2 \right]\cr
B & = & \alpha' \left({\lambda \chi \cos^2 \theta \sin^2 \theta \sinh^4 \rho  \over 
1 + \chi^2 \lambda  \cos^2 \theta \sin^2 \theta \sinh^4 \rho}\right) d \phi_1 \wedge d \phi_2 \cr
e^\phi & = & \left({1 \over 
\sqrt{1 + \chi^2 \lambda \cos^2 \theta \sin^2 \theta \sinh^4 \rho}}\right) {\lambda \over 4 \pi N} \label{deformed}
\eeq
with suitable Ramond-Ramond fields. This is a deformation of the
$AdS_5 \times S_5$ geometry (\ref{ads5xs5}) with respect to single
dimensionless parameter $\chi$. The $AdS_5 \times S_5$ geometry is
recovered in the limit $\chi \rightarrow 0$. The goal of this article
is to identify the interpretation of the deformation with respect to
$\chi$ on the field theory side of the AdS/CFT correspondence.

Precisely the deformation of this type was studied in
\cite{Beisert:2005if}, and as these authors suggested, it is quite
natural to interpret this background as being dual to a
non-commutative deformation of ${\cal N}=4$ SYM on $R \times S_3$ with
the Moyal $*$-product
\be f *g = \left.e^{{2 \pi i \chi } \left({\partial \over \partial \phi_1} {\partial \over \partial \phi_2'} - {\partial \over \partial \phi_2} {\partial \over \partial \phi_1'}\right)/2} f(\tau,\theta,\phi_1,\phi_2) g(\tau,\theta,\phi_1',\phi_2')\right|_{\phi_1 = \phi_1', \phi_2 = \phi_2'} \ . \label{moyal} \ee
This interpretation fits naturally with the established patterns seen
in other non-commutative field theories
\cite{Hashimoto:1999ut,Aharony:2000gz,Hashimoto:2002nr,Dolan:2002px,Hashimoto:2004pb,Hashimoto:2005hy}. The
naturalness of this interpretation is also echoed in
\cite{Kulaxizi:2006pp}.

There is however a problem in making this identification more precise.
The gauge/gravity dualities are motivated by the complementarity of
black D3-branes of string theory in various regimes of the t'Hooft
coupling $\lambda$ \cite{Maldacena:1997re}.  This allowed for an
explicit analysis of the physics of open string degrees of freedom,
which gave rise to a concrete realization of non-commutative dynamics
in the appropriate decoupling limit.  The $U(1) \times U(1)$ isometry
which we exploited in constructing the $\chi$ deformation is an
isometry of the near horizon $AdS_5 \times S_5$ geometry but not of
the full D3-brane geometry. This makes the direct analysis of the open
string dynamics from the world sheet point of view along the lines of
\cite{Dhokarh:2007ry} impossible.

We will show in this article that embedding into full D3 geometry is
still possible, by exploiting the underlying $SL(2,Z)$ T-duality
structure of the $(\phi_1,\phi_2)$ torus. This is the string
theoretical manifestation of the Morita equivalence in non-commutative
field theories.  To take advantage of this duality, it is useful to
restrict to the case where $\chi$ is a rational number.  Then, there
exists an $SL(2,Z)$ transformation which removes the
non-locality. Since this $SL(2,Z)$ dual is a local theory, it is the
description most suitable for exploring the deep UV behavior
\cite{Hashimoto:1999yj}. The $SL(2,Z)$ structure in fact gives rise to
a self-similar phase diagram similar to the fundamental domain of the
moduli-space of a torus. Similar structures have been shown to arise
in NCOS \cite{Chan:2001gs} and PFT \cite{Ganor:2007qh} theories as
well.  Since rational numbers are dense, this will suffice for the
purpose of identifying the field theory dual of (\ref{deformed}). In
other words, we can use the fact that the effective theory in the IR
region of the phase diagram depends smoothly on $\chi$.

Let us suppose, for sake of concreteness, that 
\be \chi = {s \over p}\ee
for relatively prime integers $p$ and $s$. Then, one can find integers
$r$ and $q$ so that
\be
\left(\begin{array}{cc} r & q \\ -s & p \end{array}\right) \in SL(2,Z)\ . \label{sl2z}
\ee
Acting on the Kahler structure $\rho'$ for the background
(\ref{deformed}) by this $SL(2,Z)$ transformation gives rise to
\be \rho'' = {r \rho' + q \over -s \rho' + p} = {q \over p} + {i \over p^2} \sqrt{\lambda}  \cos \theta \sin \theta \sinh^2 \rho \ . \ee
In other words, the supergravity background is transformed to take the form
\beq ds^2 &=&   \alpha' \sqrt{\lambda} \left[ - \cosh^2 \rho d \tau^2  + d \rho^2 + \sinh^2 \rho \left(d \theta^2 + {\sin^2\theta d \phi_1^2 +\cos^2 \theta d \phi_2^2 \over p^2}\right) 
+ d \Omega_5^2 \right]\cr
B & = & \alpha' {q \over p} d \phi_1 \wedge d \phi_2 \cr
e^\phi & = & {1 \over p^2} {\lambda \over 4 \pi N} \label{orbifold}
\eeq
where $\phi_1$ and $\phi_2$ are periodic with respect to $2 \pi$.  We
can change variables
\be \phi_i = p \tilde\phi_i, \qquad i  = 1,2 \ee
and write
\beq ds^2 &=&   \alpha' \sqrt{\lambda} \left[ - \cosh^2 \rho d \tau^2  + d \rho^2 + \sinh^2 \rho \left(d \theta^2 + \sin^2\theta d \tilde \phi_1^2 +\cos^2 \theta d \tilde \phi_2^2 \right) 
+ d \Omega_5^2 \right]\cr
B & = & \alpha' q p d \tilde \phi_1 \wedge d \tilde \phi_2 \cr
e^\phi & = & {1 \over p^2} {\lambda \over 4 \pi N} \label{orbifold2}
\eeq
with 
\be \tilde \phi_i \sim \tilde \phi_i + {2 \pi \over p}, \qquad i=1,2 \ . \ee
This solution is therefore recognizable as a $Z_p \times Z_p$ orbifold
of $AdS_5 \times S_5$ with $p N$ units of RR-flux threading the
$S_5$. This type of orbifold, acting on the $AdS_5$ sector of the
geometry, was first considered in \cite{Horowitz:2001uh}. Now, this
solution is no less easier to embed in the full D3 solution for its
dynamics to be interpreted from the open string point of view than
(\ref{deformed}), because of the orbifolding with respect to the
killing vectors
\be \xi_i = {\partial \over \partial \tilde \phi_i} , \qquad i=1,2\ .\label{xi}\ee
However, its covering space is simply $AdS_5 \times S_5$ with some
exact $B$ field.  {\it This} is easier to embed into the D3 geometry.

In order to explore the embedding into the full D3 geometry, it is
convenient to first go to the Poincare coordinate of the $AdS_5 \times
S_5$ geometry. This can be accomplished by recalling the two different
ways of parameterizing the hyperboloid
\beq 
{R \over 2u}(1+u^2(R^2+x_1^2+x_2^2+x_3^2-t^2) 
& = X_0 = & R \cosh\rho \cos \tau \cr 
R u x_1
& = X_1 = & R \sinh\rho\sin\theta\cos\tilde \phi_1\cr
R u x_2
& = X_2 = & R \sinh\rho\sin\theta\sin\tilde \phi_1 \cr
R u x_3\
& = X_3 = & R \sinh\rho\cos\theta\sin\tilde \phi_2 \cr
{R \over 2u} (1-u^2 (R^2-x_1^2-x_2^2-x_3^2+t^2)) 
& = X_4 = & R \sinh\rho\cos\theta\cos\tilde \phi_2\cr
R u t 
& = X_5 = & R \cosh\rho\sin\tau 
\eeq
satisfying $X_0^2 - X_1^2-X_2^2-X_3^2-X_4^2+X_5^2=R^2$ in $R^{2,4}$.

This implies a map between coordinates
\beq 
\tilde \phi_1 &=& \arg  \left(x_1+  i x_2  \right) \cr
\tilde \phi_2 &=& \arg \left(
{\left(-R^2-t^2+x_1^2+x_2^2+x_3^2\right)
    u^2+1  \over 2}+ i  R u^2
    x_3\right) \cr
\theta &=& \arg \left(
\sqrt{R^2 u^2
    x_3^2+\frac{\left(u^2
    \left(R^2+t^2-x_1^2-x_2^2-x_3^2\right)-1\right)^2}{4
    u^2}}+ i R u \sqrt{x_1^2+x_2^2}
\right) \cr
\tau &=& \arg \left(
{\left(R^2-t^2+x_1^2+x_2^2+x_3^2\right)
    u^2+1 \over 2} + 
i R t u^2
\right) \cr
\rho &=& \cosh^{-1} \left(\sqrt{t^2
    u^2+\frac{\left(\left(R^2-t^2+x_1^2+x_2^2+x_3^2\right)
    u^2+1\right)^2}{4 u^2 R^2}}
\right) \ . \label{map1}
\eeq
In terms of the Poincare coordinates, the supergravity background takes
on a simple form
\be ds^2 = R^2 \left( u^2 (-dt^2 + dx_1^2 +dx_2^2 + dx_3^2) + {du^2 \over u^2}+ d\Omega_5^2 \right) \label{poincare} \ee
and the $B$-field having the form
\be B = \alpha' q p {\partial \tilde \phi_1 \over \partial x_\mu}
{\partial \tilde \phi_2 \over \partial x_\nu} \, dx^\mu\wedge dx^\nu \ . \label{exactB} \ee
The fact that $dB=0$ ensures that the $AdS_5 \times S_5$ solution is
unperturbed. Suppose we rescale
\be u= {r \over R^2} \ee
which makes the metric take the form
\be ds^2 =   {r^2 \over R^2} (-dt^2 + dx_1^2 +dx_2^2 + dx_3^2) + R^2 \left({dr^2 \over r^2} +  d \Omega_5^2\right) \ .  \label{poincare2} \ee
It is then possible to extend this solution to full D3 
\be ds^2 =    \left(1+ {R^4 \over r^4}\right)^{-1/2} (-dt^2 + dx_1^2 +dx_2^2 + dx_3^2) + \left(1+ {R^4 \over r^4}\right)^{1/2} (dr^2+r^2 d \Omega_5^2) \ee
while continuing to let the $B$-field have the form (\ref{exactB})
which continues {\it not} to back react.

In the large $r$ limit, $B$ becomes
\be B = \alpha' q p \, d \tilde \phi_1 \wedge d \tilde \phi_2 \ee
where 
\be \tilde \phi_1 = \arg(x_1+ i x_2), \qquad \tilde \phi_2 = \arg( -R^2-t^2+x_1^2+x_2^2+x_3^2) \ . \ee
What this suggests is that the covering space of (\ref{orbifold2}) is
interpretable as ${\cal N}=4$ gauge theory with background field
\be F = {B \over \alpha'} = q p \, d \tilde \phi_1 \wedge d \tilde \phi_2  \label{Ffield}\ee
in the decoupling limit.  It is straight forward to verify that the
equations of motion and the Bianchi identity for the gauge fields
\be d  * F =  0 = dF \ee
are satisfied. However, since the flux is fractional, it must be
interpreted as giving rise to a 't Hooft flux \cite{Guralnik:1997sy}.

\begin{figure}
\centerline{\includegraphics[width=3in]{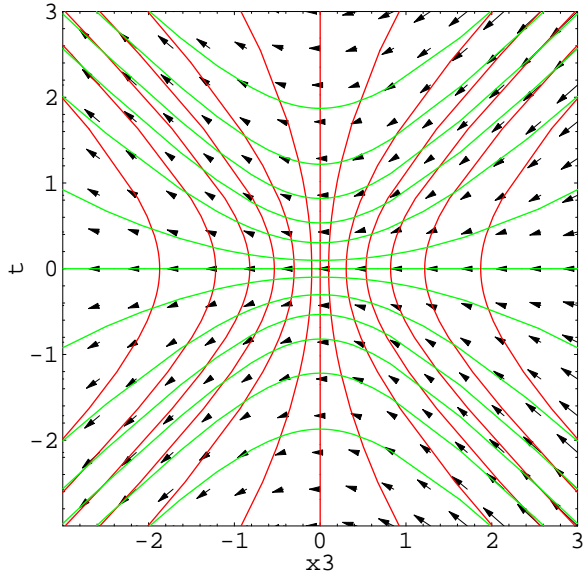}}
\caption{The contour of fixed $\tau$ (green) and fixed $\tilde \phi_2$ (red)
in the $\theta=0$ hypersurface which amounts to setting
$x_1=x_2=0$. The arrows represent the field of Killing vector
$\xi_2$. \label{figa}}
\end{figure}

Our remaining task in addressing our original motivation is to work out
the implication of (\ref{Ffield}) in identifying the field theory dual
of (\ref{deformed}). To facilitate this, it is useful to first work out
the map which relates the coordinates on the boundary of global
$AdS_5$ to the the boundary of Poincare $AdS_5$. This is achieved by
taking the large $u$ limit of (\ref{map1}) which reads
\beq 
\tilde \phi_1 &=& \arg \left(x_1+ i x_2 \right) \cr
\tilde \phi_2 &=& \arg \left(
{-R^2-t^2+x_1^2+x_2^2+x_3^2 \over 2} 
+  i R  x_3
\right) \cr
\theta &=& \arg \left(
   {\sqrt{R^2 
    x_3^2+\frac{\left(
    R^2+t^2-x_1^2-x_2^2-x_3^2\right)^2}{4
    }}} + i{R  \sqrt{x_1^2+x_2^2}}
\right) \cr
\tau &=& \arg \left(
{R^2-t^2+x_1^2+x_2^2+x_3^2\over 2
    }+ i { R t }
\right)  \ . \label{map2}
\eeq
Since we will ultimately compactify along the isometry vectors
(\ref{xi}), it would be instructive to see how these vectors are
oriented in the Poincare coordinates. We illustrate in figure
\ref{figa} the contour of fixed $\tau$ and fixed $\tilde \phi_2$ in the
$\theta=0$ hypersurface which amounts to setting $x_1=x_2=0$.

It is also useful to specify the metric for the space on which the
field theory is defined. Starting with the round metric on $R \times
S_3$
\be ds^2 = R^2 \left[d \tau^2 + d \theta^2 + \sin^2 \theta d \tilde \phi_1^2 + 
\cos^2 \theta d \tilde \phi_2^2 \right]\ee
and applying (\ref{map2}) maps this to a conformally flat  metric
\be ds^2 = f(t,x_1,x_2,x_3) (-dt^2 + dx_1^2 + dx_2^2 + dx_3^2) \ee
with
\be f(t,x_1,x_2,x_3) = \left( \frac{4 R^4}{R^4+2 \left(t^2+x_1^2+x_2^2+x_3^2\right)
    R^2+\left(-t^2+x_1^2+x_2^2+x_3^2\right)^2} 
\right) \ . \ee
Therefore, in order to interpret (\ref{orbifold2}) as a field theory
on $S^3$ with a round metric, we should start with (\ref{Ffield}) on
flat Minkowski metric, apply a conformal transformation, followed by a
diffeomorphism with respect to the map (\ref{map2}). Luckily, gauge
fields have conformal scaling dimension zero \cite{wald}. So $F$ is
invariant under conformal transformation. We therefore conclude that
(\ref{orbifold2}) is dual to ${\cal N}=4$ theory with
\be F = q p \, d \tilde \phi_1 \wedge \tilde \phi_2 \ee
with coordinates $\tilde \phi_i$ periodic under shift by $2 \pi / p$. 

To proceed further, we will view $S^3$ as $T^2$ parameterized by
$(\tilde \phi_1,\tilde \phi_2)$, fibered over an interval $I$
parameterized by $0 \le \theta \le \pi/2$. It is natural to express functions on $S^3$ in a basis 
\be f(\theta,\tilde \phi_1, \tilde \phi_2) = g(\theta) e^{i n_1 \tilde \phi_1 + i n_2 \tilde \phi_2} \ . \ee
The fact that $\tilde \phi_1$ and $\tilde \phi_2$ are periodic with
respect to shift in $2 \pi /p$ implies that $n_1$ and $n_2$ must be
integer multiples of $p$. However, in the presence of a fractional
flux \cite{tHooft:1981sz,vanBaal:1982ag}
\be \int F = q p {1 \over p^2} = {q \over p} \ , \ee
the $p\times p$ degrees of freedom in the adjoint of $SU(pN)$ splits into
adjoints of $SU(N)$ in a box whose size is larger by a factor of $p$
\cite{vanBaal:1984ar,Hashimoto:1997gm}. The non-commutative algebra of
the $p\times p$ adjoint degrees of freedom are precisely isomorphic to
the Moyal algebra with rational dimensionless non-commutativity
parameter as was shown, e.g., in
\cite{Bigatti:1998vf,Ambjorn:2000cs}. These arguments are also
reviewed in more detail in the appendix.

\begin{figure}
\includegraphics[width=\hsize]{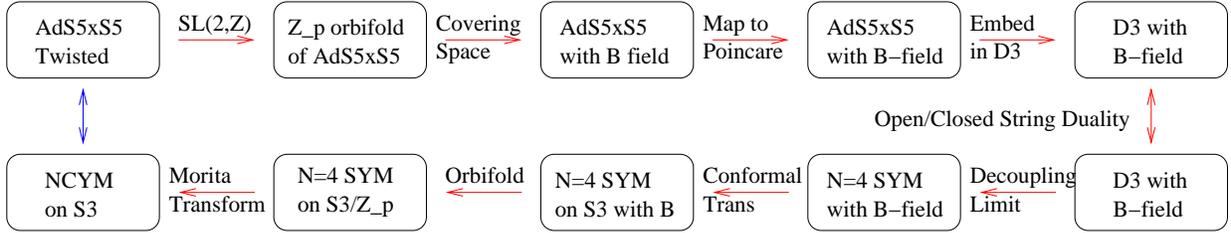}
\caption{Schematic flowchart of the duality chain, demonstrating that
the blue arrow in the far left is a consequence of the standard
open/closed string duality correspondence on the far
right.\label{figb}}
\end{figure}

Since the argument is somewhat long winded, the outline of the
argument is summarized in the flow chart diagram illustrated in figure
\ref{figb}. Our goal was to show that the Melvin twist of $AdS_5
\times S_5$ is the supergravity dual of NCSYM on $S_3$ with the
non-commutative $(\phi_1,\phi_2)$ coordinates, illustrated by a blue
arrow in figure \ref{figb}. We relied heavily on the $SL(2,Z)$
structure both on the field theory side and the supergravity side of
the correspondence, as well as the rationality of the deformation
parameter $\chi$, to reformulate the theory in terms of an orbifold of
${\cal N}=4$ theory. This allowed the duality from the open
string/closed string perspective to be made most manifest.  By
following the chain of duality back to the original description, we
derive the original duality of interest confirming
\cite{Beisert:2005if}. This is the main result of this article.

The rationality of the deformation parameter $\chi$ and subsequent
$SL(2,Z)$ transformation proved to be the powerful handle in defining
these theories at the microscopic level.  It should be possible to
formulate a microscopic formulation of Puff Field Theory along these
lines as well \cite{Haque:2008vm}.

It should be noted that strictly speaking, the deformation/orbifolding
along $\xi_i$ which we considered in this article breaks all
supersymmetries (just as in the pure Melvin case of
\cite{Hashimoto:2004pb,Hashimoto:2005hy}).  What this means is that
one expects the supergravity background to be unstable to decay, and
for the field theory side to suffer from runaway vacua. However, the
fact that the supergravity background considered in this article does
satisfy the classical equation of motion implies, as was the case for
various non-supersymmetric orbifolds \cite{Kachru:1998ys}, that the
effects of instability are subleading in $1/N$ expansion.  One could
also imagine our analysis for $\xi_1$ and $\xi_2$ in $AdS_5 \times
S_5$ which preserves some fraction of supersymmetry, such as choosing
the $\xi_1$ to be along the Hopf fiber of $S^3$, and $\xi_2$ to be
along the Hopf fiber of the $S_3$ of $SO(4) \in SO(6)$.  More specifically, 
parameterize the metric of $AdS_5 \times S_5$ by coordinates
\be ds^2 = 
R^2\left[ - \cosh^2 \rho d \tau^2  + d \rho^2 + \sinh^2 \rho d \Omega_{3(1)}^2
+ d \Omega_5^2 \right]
\ee
where
\be d\Omega_5^2 = d \alpha^2 + \cos^2\alpha d \beta^2 + \sin^2 \alpha  d  \Omega_{3(2)}^2\ee
with 
\be d \Omega_{3(i)}^2 = d \Omega_{2(i)}^2 + (d \phi_i + {\cal A}_i)^2, \qquad 
d\Omega_{2(i)}^2 = {1 \over 4} (d \theta_i^2 + \sin^2 \theta_i d \varphi_i^2) , \qquad {\cal A}_i = -{1 \over 2} (1 - \cos \theta_i) d \varphi_i\ee
and set $\xi_i = \partial_{\phi_i}$. Performing a Melvin twist by the
amount $\chi$ will give rise to a geometry
\beq  ds^2 &=& R^2\left[ - \cosh^2 \rho d \tau^2  + d \rho^2 + \sinh^2 \rho \left( d\Omega_{2(1)}^2 + {(d \phi_1 + {\cal A}_1)^2 \over (1+\chi^2 \lambda \sinh^2 \rho \sin^2 \alpha)}\right) + d \alpha^2 \right. \cr
&  & \qquad\left.+ \cos^2\alpha d \beta^2 + \sin^2 \alpha \left( d\Omega_{2(2)}^2 + {(d \phi_2 + {\cal A}_2)^2 \over (1+\chi^2 \lambda \sinh^2 \rho \sin^2 \alpha)}\right)  \right]  \eeq
which is to be interpreted as an example of a dipole field theory
\cite{Bergman:2000cw,Bergman:2001rw}. If the deformation parameter
takes on a rational value $\chi =s/p$, this geometry can be mapped,
via an $SL(2,Z)$ transformation, to $(AdS_5 / Z_p) \times (S_5/Z_p)$
geometry with torsion
\beq  ds^2 &=& R^2\left[ - \cosh^2 \rho d \tau^2  + d \rho^2 + \sinh^2 \rho \left( d\Omega_{2(1)}^2 + {1 \over p^2} (d \phi_1 + {\cal A}_1)^2\right) + d \alpha^2 \right. \cr
&  & \qquad\left.+ \cos^2\alpha d \beta^2 + \sin^2 \alpha \left( d\Omega_{2(2)}^2 + {1 \over p^2}(d \phi_2 + {\cal A}_2)^2\right)  \right]  \eeq
preserving 1/4 of the original supersymmetry and should be
stable. Other possible Killing vectors along which one can compactify
and or twist preserving some fraction of supersymmetries can be found,
e.g., in
\cite{Behrndt:1999jp,Ghosh:1999nf,FigueroaO'Farrill:2004yd,FigueroaO'Farrill:2004bz}.
Along lines similar to \cite{Hashimoto:2002nr}, many of these
constructions would constitute a laboratory for exploring issues of
string theory in time dependent backgrounds.

Finally, let us consider the thermodynamics of the twisted $U(1)
\times U(1) \in S^3$ theory from the supergravity point of view. Start
with the Schwarzschild black hole solution \cite{Witten:1998zw} \be
ds^2 = -\left({r^2 \over b^2}+1- {w_n M \over r^{n-2}}\right) dt^2 +
{dr^2 \over \left({r^2 \over b^2}+1- {w_n M \over r^{n-2}}\right)} +
r^2 d \Omega^2 \ee
where $n = 4$ for the $AdS_5$, $w_n = {16 \pi G_N \over (n-1) Vol(S^{n-1})}$,  and
\be d \Omega^2 = d \theta + \sin^2 \theta d \phi_1^2 + \cos^2 \theta d \phi_2^2 \  . \ee
The period of $t$ coordinate is given by 
\be  \beta = {1 \over T}  = {4\pi b^2 r_+ \over 4 r^2_+ +2 b^2}, \qquad {r_+^2 \over b^2}  + 1 - {w_4 M \over r_+^2}=0, \qquad r_+ = \mbox{horizon radius}\ ,    \ee
and the boundary is conformal to $S_1 \times S_3$ with periods $\beta$
and $R=b$, respectively.

One can then perform the $\chi$ deformation on this background, giving
rise to a new background
\be {ds^2  \over \alpha'} = 
 \sqrt{\lambda} \left[- \left(\cosh^2 \rho - {\mu \over \sinh^2 \rho}\right) d\tau^2 + { \cosh^2 \rho  \over 
\left(\cosh^2 \rho - {\mu \over \sinh^2 \rho}\right) }d \rho^2  + \sinh^2 \rho d \Sigma^2 \right]\label{def2}\ee
where we have changed coordinates to match the asymptotic behavior of
(\ref{ads5xs5})
\be t = R\tau, \qquad r^4 = R^4 \sinh^4 \rho=\alpha'^2 \lambda \sinh^4 \rho \ee
and
\be d \Sigma^2 = 
d \theta^2 + {\sin^2\theta d \phi_1^2 +\cos^2 \theta d \phi_2^2 \over 1 + \lambda \chi^2  \cos^2 \theta \sin^2 \theta \sinh^4 \rho}  \, 
\ee
\be \mu = {w_n M \over R^2} = \pi^4 R^4 T^4  + (\mbox{terms subleading in $1/TR$}) \ee
Just as in the undeformed case, the use of Schwarzschild black hole
solution suffers from the Hawking-Page transition at low temperatures,
but for $T > 1/R$, it follows from the standard reasoning that the
entropy
\be S(T) = {\pi^2\over 2} N^2 V T^3 \ , \ee
being proportional to the area of the horizon in the Einstein frame,
is unaffected by $\chi$.

\section*{Acknowledgements}
We would like to thank
O. Lunin and J. Simon
for discussions.
This work was supported in part by the DOE grant DE-FG02-95ER40896 and
funds from the University of Wisconsin.

\section*{Appendix}

In this appendix, we show explicitly that $U(p)$ gauge theory on a
torus of size $L \times L$ with fractional flux $q/p$ is equivalent to
a non-commutative $U(1)$ gauge theory with non-commutativity parameter
$\theta = 2 \pi s/p \times (pL)^2$ on a torus of size $pL \times
pL$. This is a standard foliation argument of non-commutative torus
\cite{Bigatti:1998vf,Ambjorn:2000cs} but we will follow the notation
and conventions of \cite{Hashimoto:1997gm}.

Consider $U(p)$ gauge theory on box size $L \times L$ with fractional
flux $q/p$. Convenient gauge is
\begin{eqnarray}
A^0_1  & = & 0 \cr
A^0_2  & = & F_0^{} x_1 I +  \frac{2 \pi}{L_2} 
{\rm Diag} (0,  1/p, \ldots, (p -1)/p) 
%F_{} & = & \frac{2 \pi}{L_1 L_2} \frac{q}{p} I
\end{eqnarray}
where
$$F_{0} = \frac{2 \pi}{L_1 L_2} \frac{q}{p}.$$
Adjoint scalars in such a background will satisfy the boundary condition
\begin{eqnarray}
\Phi (x_1 + L_1, x_2) & = &  e^{2 \pi i (x_2/L_2) T} 
V^q \Phi (x_1, x_2) V^{-q} e^{-2 \pi i (x_2/L_2) T} \nonumber \\
\Phi (x_1, x_2 + L_2) & = &    \Phi (x_1, x_2) 
\end{eqnarray}
Treating the action to the quadratic order, the plane wave solution
with this boundary condition is
$$
\delta \Phi_{m_1,m_2,r}(x_1,x_2)  =  \varphi_{m_1,m_2,r} \Lambda_{m_1,m_2,r}
e^{2 \pi i (m_1 x_1/L_1  + m_2 x_2/L_2)}
$$
where
\be m_1 \in {\bf Z}/p, \qquad m_2  \in {\bf Z}, \qquad r = 0 ... p-1 \label{quant} \ee
and
\be
\Lambda_{m_1,m_2,r} = 
{\rm Diag}\{1, \omega, \omega^2, \ldots, \omega^{p-1} \} \cdot
\left(\begin{array}{rclrcl}
   e^{-2 \pi i x_2/L_2} \!\!\!\!\!\!\!\!\!\!\!\!\!\!\!\!\!\!\!\! & & \\
   & \ddots  & \\
   & &  e^{-2 \pi i x_2/L_2}\\
&&&     1 & &  \\
&&&     & \ddots  & \\
&&&     & & 1 
\end{array}\right)
\begin{array}{l}
\left.\rule{0ex}{5.5ex}\right\}r\\
\left.\rule{0ex}{5.5ex}\right\} p-r
\end{array} \cdot V^{-r}
\ee
where $\omega = e^{2 \pi i m_1 s}$ for $qs \equiv 1\ 
({\rm mod}\; p)$.

The energy and momentum carried by these modes (see (15)-(17) of
\cite{Hashimoto:1997gm}) are
\be E^2 = k_1^2 + k_2^2, \qquad k_1 = {2 \pi m_1 \over L_1}, \qquad k_2 = {2 \pi \over L} \left(m_2 - {r \over p}\right) \ee
which in light of (\ref{quant}) is identical to that of a single degree of freedom in a box of size $pL$, rather than $p^2$ degrees of freedom in a box of size $L$.

Let us  define an algebra for the  $\varphi(m_1,m_2,r)$ that is homomorphic to the algebra of  $\Phi_{m_1,m_2,r}(x_1,x_2)$. 
In other words, we want
\be \Phi[\varphi_{k_1,k_2}(x_1,x_2) *  \varphi_{k_1',k'_2}(x_1,x_2)] = 
\Phi[\varphi_{k_1,k_2}(x_1,x_2)] \cdot \Phi [ \varphi_{k_1',k_2'}(x_1,x_2)] \ee 
where 
\be \varphi_{k_1,k_2}(x_1,x_2) =  \varphi_{k_1,k_2} e^{i k_1 x_1 + i k_2 x_2} \ee
We find 
\be \varphi_{k_1,k_2}(x_1,x_2) *  \varphi_{k_1',k'_2}(x_1,x_2) = 
e^{ik_1 \theta k'_2} (\varphi_{k_1+k'_1,k_2+k'_2}  (x_1,x_2)\ee
follows from the basic fact that 
\be \Lambda_r \Lambda_{r'} = \omega'^{-r}\Lambda_{r+r'} \ee

To see this, note that the phase factor
\be \omega'^{-r} = e^{-2 \pi i m_1' s r} = 
e^{-2 \pi i m_1' s (r - p m_2)} =
e^{{i (pL)^2 s  \over 2 \pi p}  k_1'  k_2}
\ee
from which we read off that
\be \theta = {s \over p} \cdot  {(pL)^2 \over 2 \pi} \ .   \ee
We see that this is precisely the non-commutativity parameter one
expects to find by starting with $q$ units of flux in a $U(p)$ theory
and acting by an $SL(2,Z)$ element
\be \left( \begin{array}{cc} a & b \\ c & d \end{array} \right) = 
\left( \begin{array}{cc} p & -q \\ s & r \end{array} \right) \ee
which is the inverse of (\ref{sl2z}), and which according to (1.9) of
\cite{Pioline:1999xg} maps the theory to a $U(1)$ theory with no 't
Hooft flux. The condition $qs=1 \mod p$ is precisely the $SL(2,Z)$
condition $p r + sq = 1$.

Now, this is not quite the Moyal product, but it can be shown to be isomorphic to it. Under the map
\be \varphi_{k_1,k_2}(x_1,x_2) = e^{-i k_1 \theta k_2/2} \tilde  \varphi_{k_1,k_2}(x_1,x_2) \ee
the algebra becomes
\be \tilde \varphi_{k_1,k_2}(x_1,x_2) * \tilde \varphi_{k'_1,k'_2}(x_1,x_2) = e^{i(k_1\theta k'_2 - k_2\theta k'_1)
 /2} \tilde \varphi_{k_1+k'_1,k_2+k'_2}(x_1,x_2) \ . \ee

The same argument, applied to the $T^2$ fiber of $S^3$, gives rise to
an algebra
\be \left. f(\theta,\phi_1,\phi_2) * g(\theta,\phi_1,\phi_2) = e^{2\pi i \Theta_{ij} {\partial_{\phi_i}} {\partial_{\phi'_j}}/2}  f(\theta,\phi_1,\phi_2)  g(\theta,\phi'_1,\phi'_2)  \right|_{\phi_i=\phi'_i} \ee
with
\be  \Theta_{12} = -\Theta_{21} = {s \over p} \ee
which is the  non-commutative deformation (\ref{moyal})  along $(\phi_1,\phi_2)$ coordinates of $S^3$.

\providecommand{\href}[2]{#2}\begingroup\raggedright\endgroup

\end{document}